\documentstyle[]{mn}
\newif\ifAMStwofonts

\def\etal{{\rm et al.}}

\def\simgt{\mathrel{\spose{\lower 3pt\hbox{$\sim$}}
        \raise 2.0pt\hbox{$>$}}}
\def\simlt{\mathrel{\spose{\lower 3pt\hbox{$\sim$}}
        \raise 2.0pt\hbox{$<$}}}

\ifoldfss
  \newcommand{\rmn}[1] {{\rm #1}}

  \ifCUPmtlplainloaded \else
    \NewTextAlphabet{textbfit} {cmbxti10} {}
    \NewTextAlphabet{textbfss} {cmssbx10} {}
    \NewMathAlphabet{mathbfit} {cmbxti10} {} 
    \NewMathAlphabet{mathbfss} {cmssbx10} {} 
  \fi
  \ifAMStwofonts
    \ifCUPmtlplainloaded \else
      \NewSymbolFont{upmath} {eurm10}
      \NewSymbolFont{AMSa} {msam10}
      \NewMathSymbol{\upi}     {0}{upmath}{19}
      \NewMathSymbol{\umu}     {0}{upmath}{16}
      \NewMathSymbol{\upartial}{0}{upmath}{40}
      \NewMathSymbol{\leqslant}{3}{AMSa}{36}
      \NewMathSymbol{\geqslant}{3}{AMSa}{3E}

    \fi
  \fi
\fi 

\ifnfssone
  \newmathalphabet{\mathit}
  \addtoversion{normal}{\mathit}{cmr}{m}{it}
  \addtoversion{bold}{\mathit}{cmr}{bx}{it}
  \newcommand{\rmn}[1] {\mathrm{#1}}

  \newmathalphabet{\mathbfit} 
  \addtoversion{normal}{\mathbfit}{cmr}{bx}{it}
  \addtoversion{bold}{\mathbfit}{cmr}{bx}{it}
  \newmathalphabet{\mathbfss} 
  \addtoversion{normal}{\mathbfss}{cmss}{bx}{n}
  \addtoversion{bold}{\mathbfss}{cmss}{bx}{n}
  \ifAMStwofonts
    \ifCUPmtlplainloaded \else
      \UseAMStwoboldmath
      \makeatletter
      \new@mathgroup\upmath@group
      \define@mathgroup\mv@normal\upmath@group{eur}{m}{n}
      \define@mathgroup\mv@bold\upmath@group{eur}{b}{n}
      \edef\UPM{\hexnumber\upmath@group}
      \new@mathgroup\amsa@group
      \define@mathgroup\mv@normal\amsa@group{msa}{m}{n}
      \define@mathgroup\mv@bold\amsa@group{msa}{m}{n}
      \edef\AMSa{\hexnumber\amsa@group}
      \makeatother
      \mathchardef\upi="0\UPM19
      \mathchardef\umu="0\UPM16
      \mathchardef\upartial="0\UPM40
      \mathchardef\leqslant="3\AMSa36
      \mathchardef\geqslant="3\AMSa3E
    \fi
  \fi
\fi 

\ifnfsstwo
  \newcommand{\rmn}[1] {\mathrm{#1}}

  \DeclareMathAlphabet{\mathbfit}{OT1}{cmr}{bx}{it}
  \SetMathAlphabet\mathbfit{bold}{OT1}{cmr}{bx}{it}
  \DeclareMathAlphabet{\mathbfss}{OT1}{cmss}{bx}{n}
  \SetMathAlphabet\mathbfss{bold}{OT1}{cmss}{bx}{n}
  \ifAMStwofonts
    \ifCUPmtlplainloaded \else
      \DeclareSymbolFont{UPM}{U}{eur}{m}{n}
      \SetSymbolFont{UPM}{bold}{U}{eur}{b}{n}
      \DeclareSymbolFont{AMSa}{U}{msa}{m}{n}
      \DeclareMathSymbol{\upi}{0}{UPM}{"19}
      \DeclareMathSymbol{\umu}{0}{UPM}{"16}
      \DeclareMathSymbol{\upartial}{0}{UPM}{"40}
      \DeclareMathSymbol{\leqslant}{3}{AMSa}{"36}
      \DeclareMathSymbol{\geqslant}{3}{AMSa}{"3E}
    \fi
  \fi
\fi 

\ifCUPmtlplainloaded \else
  \ifAMStwofonts \else 
    \def\upi{\pi}
    \def\umu{\mu}
    \def\upartial{\partial}
  \fi
\fi

\title[Predicting caustic crossing high magnification events in Q2237+0305]
  {Predicting caustic crossing high magnification events in Q2237+0305}
\author[J. S. B. Wyithe et al.]
  {J.~S.~B.~Wyithe$^{1,2}$, 
  R.~L.~Webster$^1$, 
  E.~L.~Turner$^2$, 
  E.~Agol$^3$ \\
  $^1$ School of Physics, The University of Melbourne, Parkville, Vic, 3052, Australia\\
  $^2$ Princeton University Observatory, Peyton Hall, Princeton, NJ 08544, USA\\
  $^3$ Physics and Astronomy Department, Johns Hopkins University, Baltimore, MD 21218\\
 Email: swyithe@astro.Princeton.edu, rwebster@physics.unimelb.edu.au, elt@astro.Princeton.edu, agol@gauss.pha.jhu.edu }
\date{Accepted. Received}
\pagerange{\pageref{firstpage}--\pageref{lastpage}}
\pubyear{1999}

\def\LaTeX{L\kern-.36em\raise.3ex\hbox{a}\kern-.15em
    T\kern-.1667em\lower.7ex\hbox{E}\kern-.125emX}

\begin{document}

\label{firstpage}

\maketitle

\begin{abstract}
The central regions of the gravitationally lensed quasar Q2237+0305 can be indirectly resolved on nano-arcsecond scales if viewed spectrophotometrically during a microlensing high magnification event (HME). Q2237+0305 is currently being monitored from the ground (eg. OGLE collaboration, Apache Point Observatory), with the goal, among others, of triggering ground and spacecraft based target of opportunity (TOO) observations of such an HME. In this work we investigate the rate of change (trigger) in image brightness that signals an imminent HME and importantly, the separation between the trigger and the event peak. In addition, we produce colour dependent model light-curves by combining high-resolution microlensing simulations with a realistic model for a thermal accretion disc source. We make hypothetical target of opportunity spectroscopic observations using our determination of the appropriate trigger as a guide. We find that if the source spectrum varies with source radius, a 3 observation TOO program should be able to observe a microlensing change in the continuum slope following a light-curve trigger with a success rate of $\ga80\%$. 

\end{abstract}


\section{Introduction}

The QSO 2237+0305, sometimes known as Huchra's lens (Huchra et al. 1985), is perhaps the most remarkable gravitational lens yet discovered. It comprises a foreground barred Sb galaxy (z=0.0394) whose nucleus is surrounded by four images of a radio-quiet QSO (z=1.695). Ground based spectroscopic observations have verified that all four images are similar QSO's at the same redshift (Adam et al. 1989). These images are separated by 0.7-0.9'' from each other and the nucleus. Broad band monitoring has shown that significant microlensing events occur (eg. Irwin et al. 1989; Corrigan et al. 1991). Since the optical depth to microlensing is of order unity at each of the image positions (eg. Kent \& Falco 1988; Schneider et al. 1988; Schmidt, Webster \& Lewis 1998), the magnification effects on the source can be considered as a network of caustics moving across the source plane. Q2237+0305 provides a unique opportunity to use microlensing events to study the geometrical structure of the background QSO for two good reasons. Firstly, the relative closeness of the lensing galaxy means that the time delay between the four quasar images is less than a day. Thus it is easy to separate microlensing events from intrinsic variations of the QSO. Also, for this QSO the time-scale of microlensing events is typically 30-50 days, reduced by a factor of 10 from typical time-scales of perhaps a year.

Detailed modelling (eg. Wambsganss \& Paczynski 1991; Fluke \& Webster 1999) shows that microlensing events resulting from the source crossing a caustic will show spectral variations which depend on the relative sizes of different emission regions at different wavelengths.
Ground based observations have confirmed differential amplification of the emission region. Lewis et al. (1998) determine the ratios of emission line equivalent widths relative to one image. They show $(i)$ the ratios remain fairly constant for one image from line to line, suggesting that the sizes of the emission regions for the lines are not greatly different, $(ii)$ that the ratios vary from image to image for a single epoch by a factor of $\sim 2.5$, and $(iii)$ that the ratio for a single image varies as a function of time, i.e. as a result of a microlensing event. These results are consistent with earlier results of Fillipenko (1989) who measured a $\sim25\%$ difference in the width of the MgII lines between the A and B images.   

A QSO is generally described as a continuum source surrounded by small gas clouds which are photo-ionised by the central source. The size of the central continuum source in a QSO is expected to be at least ten times the gravitational radius ($r_g$) of the black hole thought to power the source: $r_g\sim10^{13}(M_{BH}/10^8\,M_{\odot})\,cm$. This size is corroborated by the length scale, $ct\sim10^{14}t_{hours}\,cm$, associated with X-ray and optical changes in the continuum emission.

Reverberation mapping provides one technique for determining the geometric structure of a QSO, although the technique has mostly been applied to nearby Seyferts. Substantial monitoring programs have already confirmed a picture where the continuum region and the region emitting the highest ionisation lines vary faster and are therefore smaller than the resolution scale of the reverberation technique. Typical sampling time-scales are several days, which corresponds to $2.5\times 10^{15}\,t_{days}\,cm$. 

Microlensing allows us to probe length scales which are an order of magnitude smaller than this, and which will not be resolvable by any other technique in the foreseeable future. Already variability amplitude (Wambsganss, Paczynski \& Schneider 1990) and time (Wyithe, Webster, Turner \& Mortlock 2000 (hereafter WWTM00))  constraints from the observed microlensing event limit the R-band continuum emission region to be less than $2\times10^{15}\,h_{75}^{-\frac{1}{2}}\,cm$. This size limit has been used by Rauch and Blandford (1991) to argue that the continuum emission cannot result from a thermal accretion disc, although Jarozynski et al. (1992) dispute this result. Spectrophotometric observations during a HME could resolve the issue. Although simulation of caustic crossings using realistic accretion disk models is the most straightforward approach theoretically and the least demanding of observational data quality, it is an intrinsically model dependent mode of investigation. A different approach is to invert the light curve observed to directly obtain a surface brightness profile of the source (eg. Grieger et al. 1998; Grieger et al. 1991; Agol \& Krolik 1999; Mineshige \& Yonehara 1999).

We discuss microlensing theory related to a proposed experiment to measure spectroscopic change during an HME. The experiment is a target of opportunity program with Hubble Space Telescope, where the observations would be triggered by the observation of the onset of a caustic crossing from ground based monitoring of the QSO. The goal is to observe a caustic passing across the continuum region, magnifying different length-scales as a function of time. The wave-lengths chosen will cover the Ly Limit, Ly$\beta/$OVI, Ly$\alpha$, NV OI, SiIV/OIV, CIV, HeII and MgII emission lines and a number of associated absorption lines, covering more than 3700$A$ in the rest frame of the QSO. The long wavelength baseline will enable determination of whether the continuum emission at all wavelengths is emitted from the same geometric regions or whether some regions are bluer.

This paper discusses the light-curve derivatives (triggers) in monitoring data that signal an imminent HME. The aim is to determine a reliable function to predicting the onset of an HME from monitoring data in real time.
 We then discuss one potential outcome of the TOO experiment; the spectral change of a thermal continuum during an HME. Sample triggering scenarios are applied to model microlensed light curves produced for a thermal accretion disc, and the probability of success of an example target of opportunity program is discussed.

\section{The Microlensing model}
\label{model}

 Throughout the paper, standard notation for gravitational lensing is used. The Einstein radius of a 1$M_{\odot} $ star in the source plane is denoted by $\eta_{0} $. The normalised shear is denoted by $\gamma$, and the convergence or optical depth by $\kappa$. The basic model for microlensing at high optical depth was first described by Kayser, Refsdal \& Stabell (1986). It comprised a disc of point masses having a size such that a large fraction ($>99\%$) of macroimage flux is recovered (Katz, Balbus \& Paczynski 1986; Lewis \& Irwin 1995). To construct a microlensed light-curve we use the contouring technique of Lewis et al. (1993) and Witt (1993). For the microlensing models of Q2237+0305 presented in the current work we assume the macro-parameters for the lensing galaxy calculated by Schmidt, Webster \& Lewis (1998). The standard notation introduced by Yee (1988) is used to describe these images.
 Where required a cosmology having $\Omega=1$ with $H_{o}=75\,km\,sec^{-1}$ is assumed. 

We describe the microlensing rate using the effective transverse velocity which is defined as the transverse velocity that produces a microlensing rate from a static model equal to that of the observed light-curve (Wyithe, Webster \& Turner 1999 (hereafter WWT99)). The effective transverse velocity therefore describes the microlensing rate due to the combination of the effects of a galactic transverse velocity and proper motion of microlenses. Our calculation of the light-curve derivatives that precede an HME assume that the effective transverse velocity accurately describes not only the distribution of light-curve derivatives during an HME (Wyithe, Webster \& Turner 2000a (hereafter WWT00a), but also the distribution of orientations between the caustic and source trajectory, and hence the event duration.

We have previously obtained the following normalised probability distributions. These were obtained under the assumption that the source size $S\ll\eta_o$. Evidence in favour of this assumption was presented in Wyithe, Webster \& Turner (2000c).

\noindent $i)$ $p_{s}(S|\langle m \rangle,v_{eff})$, the probability that the continuum source diameter is between $S$ and $S + {\rmn d}S$ given a mean microlens mass $\langle m \rangle$ and an effective galactic transverse velocity $v_{eff}$ (WWTM00).

\noindent $ii)$ $p_{v}(v_{eff}|\langle m \rangle)$ the probability that the effective galactic transverse velocity is between $v_{eff}$ and $v_{eff}+{\rmn d} v_{eff}$ given a mean microlens mass $\langle m \rangle$ (WWT99). 

\noindent $iii)$ $p_{m}(\langle m \rangle)$, the probability that the mean microlens mass is between $\langle m \rangle$ and $\langle m \rangle + {\rmn d} \langle m \rangle$ (Wyithe, Webster \& Turner 2000b).

$p_{v}(v_{eff}|\langle m \rangle)$ and $p_{m}(\langle m \rangle)$ were computed using flat ($p(V_{tran})\propto dV_{tran}$), and logarithmic ($p(V_{tran})\propto \frac{dV_{tran}}{V_{tran}}$) assumptions for the Bayesian prior for galactic transverse velocity ($V_{tran}$). $p_{v}(v_{eff}|\langle m \rangle)$ was found to be insensitive to the prior assumed, however $p_{m}(\langle m \rangle)$ showed some dependence. In the remainder of this paper we use $p_{m}(\langle m \rangle)$ calculated using the assumption of a logarithmic prior. We note that the assumption of the flat prior raises the average light-curve derivative by a few percent.

 The functions $p_{s}(S|\langle m \rangle,v_{eff})$, $p_{v}(v_{eff}|\langle m \rangle)$, $p_{m}(\langle m \rangle)$ and the HME statistics presented in this paper were computed for the following assumptions of smooth matter density, photometric error, and direction of the galactic transverse velocity.
Two models are considered for the distribution of microlenses, one with no continuously distributed matter, and one where smooth matter contributes 50\% of the surface mass density. Two orientations were chosen for the transverse velocity with respect to the galaxy, having source trajectories parallel to the A$-$B and C$-$D axes. The two orientations bracket the range of possibilities, and because the images are positioned approximately orthogonally with respect to the galactic centre correspond to shear values of $\gamma_{A},\gamma_{B}<0,\gamma_{C},\gamma_{D}>0$ and $\gamma_{A},\gamma_{B}<0,\gamma_{C},\gamma_{D}>0$ respectively. Photometric error was simulated by perturbing the model light-curve by an amount distributed randomly from a Gaussian of halfwidth $\sigma$. The simulations used two different estimates of the error in the photometric magnitudes. In the first case a small error was assumed (SE). For images A and B, $\sigma_{SE}$=0.01 mag, and for images C and D $\sigma_{SE}$=0.02 mag. In the second case, a larger error was assumed (LE). For images A and B, $\sigma_{LE}$=0.02 mag and for images C and D $\sigma_{LE}$=0.04 mag. The random component of observational error quoted in Irwin et al. (1989) was 0.02 mag. 

 Both the microlensing rate due to a transverse velocity (eg. Witt, Kayser \& Refsdal 1993), as well as the corresponding rate due to proper motions (WWT00a) are not functions of the details of the microlens mass distribution, but rather are only dependent on the mean microlens mass. We therefore limit our attention to models in which all the microlenses have the same mass since the results obtained will be applicable to other models with different forms for the mass function.

\section{Intrinsic variation of Q2237+0305}
\label{intrinsic}

\begin{figure*}
\vspace{140mm}
\includegraphics{fig1.epsi}
\caption{\label{intrinsic}Exclusion plots for the light-curve variance and power-spectrum index $\gamma_I$ for intrinsic variability in Q2237+0305. Limits are imposed by consideration of a) light-curve variance, b) light-curve variance about local mean, c) histogram of light-curve derivatives, and d) autocorrelation time-scales.}
\end{figure*}

Intrinsic variation of the quasar in the 2237+0305 system cannot be directly measured since one cannot be sure how much variation is due to microlensing and how much is intrinsic to the source. 
The afore-mentioned analyses of microlensing in Q2237+0305 to determine $p_s$, $p_v$ and $p_m$ have utilised difference light-curves. This is practical for Q2237+0305 since the relative time-delays between images ($<$1-day (eg Schneider et al. 1988)) are small compared with typical sampling time-scales (weeks or months). However our current goal is to determine whether a particular observed light-curve derivative signals a forthcoming HME. Since an observed (single image) light-curve derivative depends on the super-position of microlensing and intrinsic source variation, the calculation of an event trigger requires the inclusion of a statistical description of the intrinsic source variation.

Giveon et al. (1999) analysed optical variability in the sample of Palomar-Green quasars. They find that over time-scales between 100 and 1000 days the sources have a power spectra that is of a power-law form, $P_{\nu}\propto\nu^{\gamma_I}$ with $\gamma_I\sim-2$. We model intrinsic variability for Q2237+0305 using a power-spectrum of power-law form. Since long duration intrinsic variability may have been observed in Q2237+0305 ($\O$stensen et al. 1995), and our current focus is on the potential for rapid intrinsic variability to cause a false trigger, we calculate model intrinsic light-curves with power spectra $P_{\nu}\propto\nu^{\gamma_I}$ over a broad range of time-scales: between 1 and 1000 days. We consider a 2-D parameter space of intrinsic light-curve variance ($\sigma_I^2$) and power-spectrum index $\gamma_I$, and look for regions of this space whose values, when combined with microlensing models produce variability statistics that are inconsistent with observation. We first describe a method for computing the likely-hood (given assumed values for $\sigma_I$ and $\gamma_I$) for a statistic $f$ defined such that larger values of $f$ imply a large intrinsic contribution to variability. The likely-hood is then compared to the observed value $f(obs)$ for that statistic. We place limits on the values for $\sigma_I^2$ and $\gamma_I$ using 4 different, but related light-curve statistics $f$. 

 Pairs of single image light-curves can be combined to produce both difference and additive light-curves. While intrinsic and microlensed variability cannot be directly distinguished, intrinsic variation is only present in the latter. In addition, microlensed variability in different images is independent. Variability statistics for difference and additive light-curves in the absence of intrinsic source variability should therefore be identical (on average). Thus comparison with the data of variability statistics for these two curves, averaged over 6 image pairs (only three are independent) provides a means of discrimination between different models for intrinsic source variability.

The statistic $f$ is calculated for additive ($f_A$) and difference ($f_D$) light-curves. Light-curves that include both intrinsic and microlensed fluctuation exhibit higher variability in the additive light-curve ($f_A>f_D$). By comparing $f_{A-D}=f_A-f_D$ calculated from the monitoring data with values obtained from mock observations for different assumptions for the rate and amplitude of intrinsic variability ($\sigma_I^2$, $\gamma_I$), we compute the probability of obtaining the observed $f_{A-D}(obs)$ as a function of $\sigma_I$ and $\gamma_I$. Microlensing statistics vary with effective transverse velocity and sampling rate. We have therefore computed this probability as a function of effective transverse velocity, using mock data sets computed from light-curves with the observed sampling rate:
\begin{equation}
\label{prob_var1}
P(f_{A-D}<f_{A-D}(obs)|\sigma_I,\gamma_I,v_{eff},\langle m\rangle)
\end{equation}
We present calculations using mock observations that include components of systematic ($\sigma_{sys}=0.04$) and random ($\sigma_{SE}$) photometric uncertainty. To obtain $P(f_{A-D}<f_{A-D}(obs)|\sigma_I,\gamma_I)$ Eqn. \ref{prob_var1} is convolved with $p_v(v_{eff}|\langle m\rangle)$:
\begin{eqnarray}
\nonumber
&&P(f_{A-D}<f_{A-D}(obs)|\sigma_I,\gamma_I) \\
&&\hspace{0mm}=\int\,P(f_{A-D}<f_{A-D}(obs)|\sigma_I,\gamma_I,v_{eff},\langle m\rangle)\\\nonumber
&&\hspace{10mm} \times p_v(v_{eff}|\langle m\rangle)\,dv_{eff}
\end{eqnarray}
Thus, the probability that $\gamma_I<\gamma_o$ and $\sigma_{I}<\sigma_{I,o}$ is 
\begin{eqnarray}
\label{prob_var2} 
\nonumber
&&P(\gamma_I<\gamma_o\,\,\&\,\,\sigma_{I}<\sigma_{I,o}|f_{A-D}(obs))\\
&&\hspace{-7mm}=\int\,P(f_{A-D}<f_{A-D}(obs)|\sigma_I,\gamma_I)\,p_{prior}(\gamma_I,\sigma_I)d\gamma_I\,d\sigma_I
\end{eqnarray}
We assume that $p_{prior}(\gamma_I,\sigma_I)\propto 1$.

Fig. \ref{intrinsic} shows plots of the 1, 5, 10, 90, 95 and 99\% contours for Eqn. \ref{prob_var2} calculated for 4 different variability statistics $f$, which we describe below. Statistics $i)$, $ii)$ and $iv)$ were calculated using the full set of observations. Statistic $iii)$ employs a derivative analysis and was therefore calculated from a modified data set in which some adjacent points were averaged to reduce noise. We note that this reduces the potential of statistic $iii)$ to measure intrinsic variability of high frequency. 

We include recent data from the OGLE collaboration (Wozniak et al. 2000a,b; OGLE web-page), which has a much denser sampling rate ($\sim$1 point per week). This increases the sensitivity to measuring high frequency intrinsic variability. While the OGLE data is in V-band, data taken prior to 1996 is in the r and R-bands. Microlensing is colour dependent when the source is near a caustic (Wambsganss \& Paczynski 1991). In Q2237+0305 $(S\ll\eta_o)$, this is only the case a small fraction of the time and we therefore assume that the microlensing statistics are consistent between V and R-bands. Any colour dependence of intrinsic variability is expressed in the additive light-curve. However, Giveon et al. (1999) find that the rms amplitudes for the sample of Palomar-Green quasars for B and R bands differ by only $\sim15\%$. We therefore assume that the power-spectrum parameters we obtain apply equally well to intrinsic source variability in both the R and V-bands, and use all the available data points without application of a colour correction.

\noindent $i)$ Light-curve variances: In the absence of intrinsic variability, the variance of additive ($V_A^2$) and difference ($V_D^2$) light-curves are identical. However, the variance of light-curves that include both intrinsic ($\sigma_I^2$) and microlensed fluctuation obey $V_A^2 \sim V_D^2+2\sigma_I^2$. Therefore the statistic 
\begin{equation}
f\equiv V_A-V_D
\end{equation}
can be used to place limits on intrinsic variability. From Fig. \ref{intrinsic}a we find that at the 95\% level the variance of intrinsic variation is limited to be $\sigma_I<0.28$ (for all $\gamma_I$ considered). The light-curve variance does not discriminate between different values of $\gamma_I$.

\noindent $ii)$ Local light-curve variance: To probe the shorter frequency intrinsic variation, we have modified statistic $i)$ in the following way. Each light-curve is defined by $N_{obs}$ observations, with $t_{i}$ and $\Delta M_{i}$ representing the time of observation and light-curve magnitude respectively. For each light-curve we calculate the variance about a $local$ mean defined by:
\begin{equation}
\langle \Delta M\rangle_{local}(t) = \frac{\sum_{i=0}^{N_{obs}}\Delta M_i\,\Pi(t_i-t)}{\sum_{i=0}^{N_{obs}}\Pi(t_i-t)}
\end{equation}
where 
\begin{equation}
\Pi(t_i-t)= \begin{array}{cc}
         1 & \rmn{for} -185<t_i-t<185\,\rmn{days}  \\
         0 & \rmn{otherwise}
            \end{array}. 
\end{equation}
The variance $V^2(local)=\langle (\Delta M_i-\langle \Delta M\rangle_{local}(t))^2\rangle$ measures fluctuations in the light-curve with durations $\la$ 1 year. Fig. \ref{intrinsic}b is a plot of 90, 95 and 99\% contours from Eqn. \ref{prob_var2} calculated for 
\begin{equation}
f\equiv V_A(local)-V_D(local)
\end{equation}
Tighter limits are imposed on the amplitudes of more rapid variations. As an example, the combination of $\gamma_I=-1$ and $\sigma_I=0.2$ is ruled out at the 95\% level.

\noindent $iii)$ Light-curve derivatives: Light-curve derivatives can be used to study quasar variability. Histograms of light-curve derivatives that result from difference and additive light-curves have a shape that differs more where the intrinsic variability is higher. $R$ is defined as the value required to multiply the difference light-curve derivatives such that the KS difference between the resulting histogram and the additive light-curve histogram is minimised.
We have computed Eqn. \ref{prob_var1}, with
\begin{equation}
f\equiv R,
\end{equation}
 and plotted the 5, 10, 90, 95 and 99\% contours in Fig \ref{intrinsic}c. For intrinsic variability having a variance that is small with respect to that of microlensing the statistic $R$ demonstrates similar behaviour to $V^2(local)$. However, since the distribution of microlensed light-curve derivatives has a different shape to the distribution resulting from a power spectrum, the minimised KS difference is large if the intrinsic variance is high. This method is therefore inapplicable if the intrinsic variability is comparable to the microlensed variability. On the other hand, if intrinsic variability is assumed to have a small amplitude with respect to microlensed variability, $R$ provides the tightest limits on $\sigma_I$. 

\noindent $iv)$ Autocorrelation time-scales: Variability can also be studied through computation of the autocorrelation function (ACF):
\begin{equation}
ACF(\tau)\equiv\langle(\Delta M(t)-\langle M\rangle).(\Delta M(t+\tau)-\langle M\rangle)\rangle.
\end{equation}
 We estimate the ACF by calculating the z-transformed discrete autocorrelation function (Alexander 1997). The width of the ACF peak (which we take to be the zero crossing time) can be used to define a correlation time-scale ($T$). Significant intrinsic variability will effect $T$. The 4th statistic is the difference between the correlation time-scale for the additive ($T_A$) and difference ($T_D$) light-curves
\begin{equation}
f\equiv T_A-T_D.
\end{equation}
The 50\% and 90\% contours of Eqn. \ref{prob_var1} are shown in Fig \ref{intrinsic}d. The autocorrelation time-scale suggests that the amplitude of intrinsic variability is small. However, it does not place useful limits on $\sigma_I$ or $\gamma_I$ since due to its larger variance, the auto-correlation time-scale is dominated by microlensing variability.

These results suggest that the intrinsic variability has small amplitudes and long durations. 

\section{Determination of a trigger for TOO observations of an HME}

\begin{figure}
\vspace{75mm}
\includegraphics{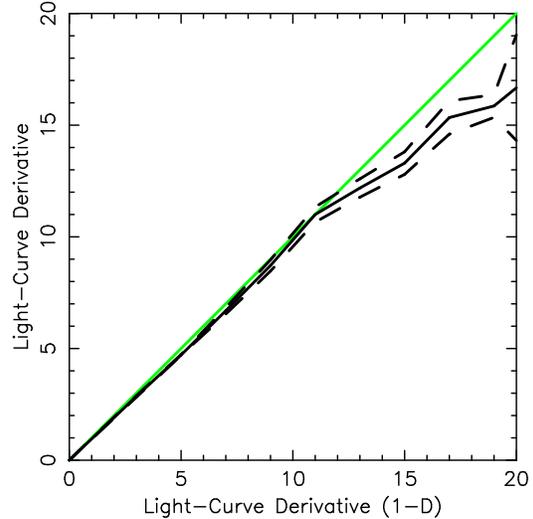}
\caption{\label{deriv_check}Correlation plot of light-curve derivatives for a model light-curve of image A. The plot shows the cases of a full 2-dimensional convolution of the source intensity profile with the microlensing magnification grid vs. the single source-line approximation. The solid and dashed lines show the mean and $\pm1\sigma$ correlations. The light line is the line of equivalence.}  
\end{figure}

There are many properties of the Q2237+0305 light-curves that could be used as a trigger, ie. features in the observations that signal the onset of an event and provide information on when that event occur. These range from a very simple property such as the light-curve derivative, to a complex procedure involving fitting functional forms or even sections of model light-curve to the monitoring data. In the instance of perfect monitoring data, a simple trigger would provide ambiguous results, while more complex procedures will be more accurate. However, given the finite accuracy of monitoring data, simple procedures will provide more definite predictions since photometric noise is amplified in higher derivatives. We therefore investigate the light-curve derivative as a tool to predict caustic crossings.

There are two important quantities: $i)$ What is the light curve derivative that signals the onset of a caustic crossing, and $ii)$ given that derivative, what is the separation between the current observation and the event peak. We calculate probability functions for the separation between the observation of a rapid light-curve rise and the subsequent HME. Note that while it would still be interesting, a drop in observed flux indicates that the HME has already peaked.

We define a trigger as a 3-point light-curve derivative that lies above a value $T$. Due to the sparse sampling of monitoring campaigns, higher order numerical derivatives may have a baseline that is large with respect to the time between the observation and a subsequent event. To investigate the information on future HMEs contained in trigger using model light-curves $M(t)$, we step along the light-curve, looking for a derivative $\dot{M}>T$, and examine the subsequent light-curve for an HME. Upon finding a trigger, our algorithm locates the next event and resumes the search from there. Thus each event is only found once during a pass along the light-curve.

In the calculations that follow, caustic crossing HMEs are defined to occur at the time when the caustic crosses the centre of the source. We note that this is not the same as the time of the maximum event amplitude, but it is the more relevant time with respect to spectroscopic observations. However these points differ by only a few days. A cusp event is defined as occurring when the light-curve reaches its local maximum.

 We have assumed a Gaussian source profile. Because it is impractical to compute the full integrations of extended sources over the range of source size and microlens mass considered, we have approximated the extended source light curve by an integral along a single source line (eg. Witt \& Mao 1994). To check the validity of this approximation we have computed light curves for a Gaussian source using both the 1-D approximation and a full 2-D integration over a 50$\times$50 grid as described in Wyithe \& Webster (1999). The light-curves were computed for a 2$\times 10^{15}\,cm$ source, which is the upper limit of source size for Q2237+0305 (Wambsganss, Paczynski \& Schneider 1990; WWTM00), and microlenses of $1$M$_{\odot}$. Fig. \ref{deriv_check} is a plot of light-curve derivatives calculated from the 1-D approximation vs. those calculated from the 2-D convolution (the mean and $\pm 1\sigma$ lines are shown). The plot shows that there is some scatter, but that the approximation is a reasonable one at all gradients. The plot demonstrates the validity of the approximation in the case of the upper limit for source size (where it is least valid). The relationship is tighter for smaller sources.

In Sec. \ref{general} (limiting our attention to images A and C) we discuss the calculation of a function that describes the onset of a HME based on observed light-curve derivatives. Secs. \ref{sec_systematics}, \ref{sec_intrinsic} and \ref{sec_sourcesize} discuss the systematic effects of unknowns on the accuracy of the function obtained, and Sec. \ref{sec_sampling} discusses the appropriate sampling rate for monitoring data. In Secs. \ref{sec_classes}, \ref{selecting} and \ref{sec_alternative} we present triggering functions specific to different classes of HME, discuss the optimal trigger and briefly discuss a method for determining the function appropriate for an isolated light-curve derivative.

\subsection{A general function for HME triggering}
\label{general}

\begin{figure*}
\vspace{160mm}
\includegraphics{fig3a.epsi}
\includegraphics{fig3b.epsi}
\caption{\label{contour_trigger_AC}$F_T(P|T)$ (left) and $\partial F_T(P|T)/\partial P$ (right). The microlensing parameters correspond to those of image A (top) and C (bottom).}
\end{figure*}

We would like to know the trigger that will best anticipate an oncoming HME. We consider a grid of triggers $T$ and periods of time $P$ following the trigger. For each trigger $T$ we calculate the probability of observing a HME within each period $P$ for a collection of assumed effective galactic transverse velocities, source sizes and mean microlens masses:
\begin{equation} 
F_T\left(P|T,S,\langle m\rangle,v_{eff}\right).
\end{equation}
This function is calculated for HMEs located on 1000$\langle m\rangle\eta_o$ of light-curve, and is evaluated for combinations of 50 values for $v_{eff}$~($0\,km\,sec^{-1}<v_{eff}<2000\,km\,sec^{-1}$, 20 values for $\langle m\rangle$~(0.001$M_{\odot}<\langle m\rangle<100M_{\odot}$) and 20 values for $S$~($10^{11}cm<S<10^{16}cm$).  
The probability for observing a HME in the period $P$ following a trigger $T$ can then be obtained:
\begin{eqnarray}
\nonumber
&&\hspace{-7mm}F_T\left(P|T\right)=\\
\nonumber&&\hspace{-3mm}\int dm\int dS\int dv_{eff}\,\,\left( p_{s}\left(S|\langle m\rangle,v_{eff}\right)\,p_{m}\left(\langle m\rangle\right)\right.\\
&&\hspace{10mm}\times\left.p_{v}\left(v_{eff}|\langle m\rangle\right)F_T\left(P|T,S,\langle m\rangle,v_{eff}\right)\right)
\end{eqnarray}

The upper and lower plots of Fig. \ref{contour_trigger_AC} show contours of $F_T(P|T)$ (left) and $\partial F_T(P|T)/\partial P$ (right) for images A and C respectively. The sampling rate was 1 point per 2 days. The functions quantify the intuitive notion that steeper observed light-curve gradients imply a more imminent event. In addition, the plots show that the separation of the trigger and event peak is less sensitive to trigger size for larger triggers. There is little difference between the functions $F_T(P|T)$ calculated for images A and C.

\subsection{The effect of model assumptions on the triggering function}
\label{sec_systematics}
\begin{figure}
\vspace{145mm}
\includegraphics{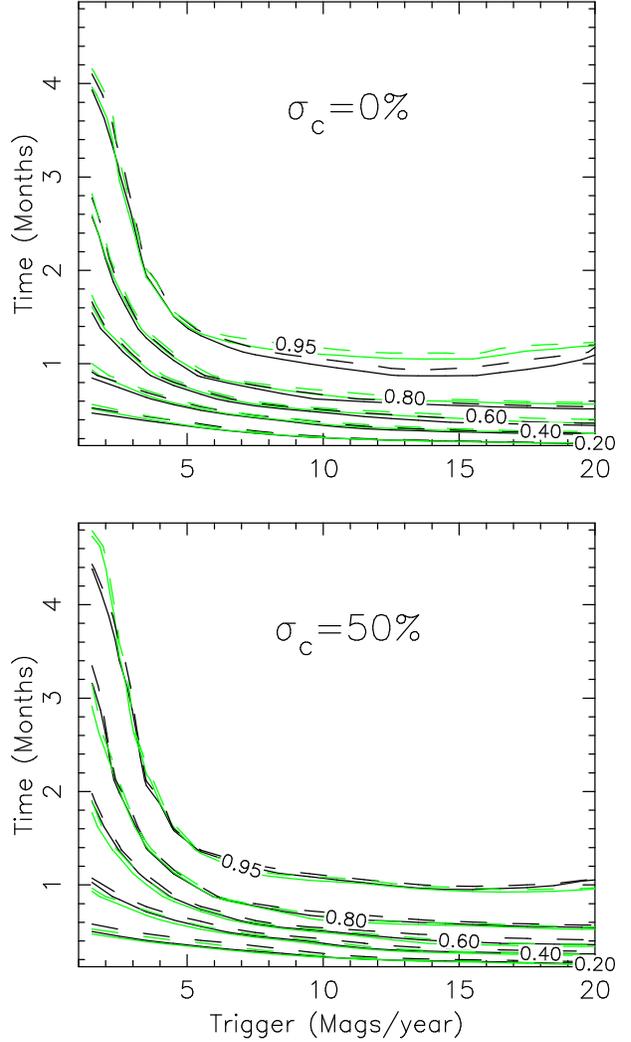}
\caption{\label{contour_sys_0}$F_T(P|T)$ (left) and $\partial F_T(P|T)/\partial P$ (right) for combinations of the different assumptions used in the calculations of $p_v$, $p_s$ and $p_m$. The solid and dashed lines correspond to small (SE) and large (LE) errors respectively. The light and dark lines correspond to $\gamma_A>0$ and $\gamma_A<0$. The microlensing parameters correspond to those of image A.}
\end{figure}

We have investigated the systematic dependence of $F_T(P|T)$ on our different model assumptions.
Fig. \ref{contour_sys_0} shows $F_T(P|T)$ for combinations of shear direction, smooth matter percentage and the photometric error in published monitoring data used to estimate $p_m$, $p_s$ and $p_v$. Plots are shown for image A, and were computed with a sampling rate of 1 point per 2 days. These plots show that the triggering function is quite insensitive to these systematic uncertainties. This is not surprising since the effective transverse velocity was determined for these different cases from a derivative analysis. In the remainder of this paper we consider only the case where $\gamma_A>0$, $\sigma_C=$0\% and the photometric errors are small (SE).

\subsection{The effect of intrinsic source variation on the triggering function}
\label{sec_intrinsic} 
\begin{figure}
\vspace{145mm}
\includegraphics{fig5.epsi}
\caption{\label{intrinsic_trigger}Contour plots for $F_T(P|T)$ computed from model light-curves which include an intrinsic component of variability with a power-spectrum having a power-law form with index $\gamma_I=-2$ and a variance $\sigma_I=0.2$ (top) and  $\sigma_I=0.4$ (bottom). The dashed contours in each case show $F_T(P|T)$ in the absence of intrinsic variability. There is no smooth matter in these models. The microlensing parameters correspond to those of image A ($\gamma_A>0$), and the photometric errors were assumed to be small (SE).}
\end{figure}

The gradient of a light-curve is influenced by intrinsic variation of the source. We include intrinsic source variation in our simulations according to a power-law shaped power-spectrum with an index of $\gamma=-2$ (following the findings of Giveon et al. (1999)) and periods between 1 and 1000 days. Sec. \ref{intrinsic} suggests that an intrinsic average light-curve variance of $\sigma_I=0.2$ is higher than expected and that $\sigma_I=0.4$ is very unlikely. Fig. \ref{intrinsic_trigger} shows $F_T(P|T)$ calculated with intrinsic fluctuation included for the cases of $\sigma_I=0.2$ (top), and $\sigma_I=0.4$ (bottom). Contours for $F_T(P|T)$ with no intrinsic variation are included for comparison (dashed lines). The former has minimal effect, while a large intrinsic variance causes false triggers (no HME) at lower trigger values. We conclude that intrinsic variation should not have an important observable effect on triggering of HME observations in Q2237+0305.

\subsection{The effect of an incorrect source size on the triggering function}
\label{sec_sourcesize}
\begin{figure*}
\vspace{60mm}
\includegraphics{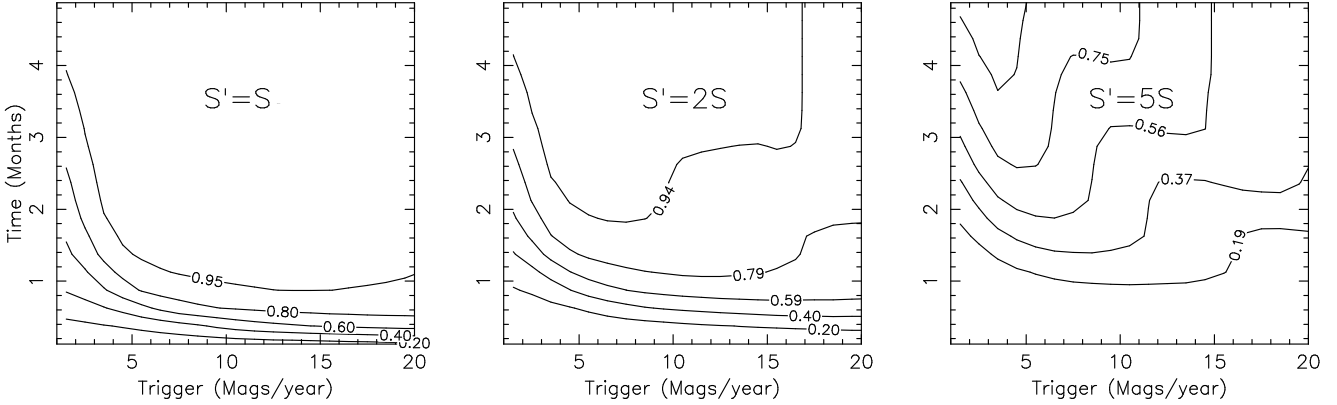}
\caption{\label{contour_size}Contour plots for $F_T(P|T)$ computed from model light-curves having a size distributed according to our best estimate $S$ (left), $S'=2S$ and $S'=5S$. There is no smooth matter in these models, small photometric errors (SE) and $\gamma_A,\gamma_B>0$ are assumed.}
\end{figure*}

The determination of probability for the quantity $v_{eff}\sqrt{\langle m\rangle}$ from the Q2237+0305 monitoring data is quite robust. However, the source size estimate is more uncertain, having been derived from a single poorly sampled HME. The small number of observations describing the 1988 peak (Irwin et al. 1989; Corrigan et al. 1991) introduces the potential for a systematic error in the estimate of source size equal to the ratio of true event length divided by the inferred event length of 52 days (twice the estimated rise time). This can be compared to the $\sim$100 day separation of the two points that provide an upper-bound to the event duration. This systematic error in the estimate of source size is therefore smaller than a factor of 2. Fig. \ref{contour_size} shows the effect (on $F$) of replacing $p_{s}\left(S|\langle m\rangle,v_{eff}\right)$ with 
\begin{equation}
p'_{s}\left(S'|\langle m\rangle,v_{eff}\right)dS'=\frac{1}{2}p_{s}\left(2S|\langle m\rangle,v_{eff}\right)dS
\end{equation}
(centre) and 
\begin{equation}
p'_{s}\left(S'|\langle m\rangle,v_{eff}\right)dS'=\frac{1}{5}p_{s}\left(5S|\langle m\rangle,v_{eff}\right)dS
\end{equation}
(right). Contours of the resulting functions $F_T(P|T)$ are shown together with $F_T(P|T)$ computed from our best estimate of $p_{s}$ for comparison (left hand figure).
Inspection of Fig. \ref{contour_size} shows that if the systematic error in the estimation of source size is the worst case of a factor of 2, then the forecasting of events will be effected if the trigger is greater than $\sim 5$ magnitudes per year. However, the forecasting of events is far more uncertain for all triggers if the estimate of source size is incorrect by a factor of 5. While the larger source produces flatter peaks, longer events and hence a smaller maximum derivative, the large scale derivatives occur earlier due to parts of the source having moved into close proximity of the caustic sooner. Hence the larger assumption for source size predicts a larger separation between trigger and corresponding event peak.

\subsection{What is the appropriate sampling rate for monitoring of Q2237+0305?}
\label{sec_sampling}
\begin{figure}
\vspace{75mm}
\includegraphics{fig7.epsi}
\caption{\label{contour_sampling}Contour plots for $F_T(P|T)$ computed from model light-curves with sampling rates of 2, 8, 16 and 32 days. There is no smooth matter in these models, small photometric errors (SE) and $\gamma_A,\gamma_B>0$ are assumed.}
\end{figure}

It would be ideal to monitor the Q2237+0305 daily. However as this is not always possible we investigate the success of predicting the onset of a caustic crossing using different simulated sampling rates. The light-curve derivatives have been calculated with spacings of 2, 8, 16 and 32 days to mimic a trigger obtained from regular monitoring. Fig. \ref{contour_sampling} shows plots of the functions $F_{T}(P|T)$ computed with the above sampling rates. Here we have defined the delay of the event following the trigger with respect to the central (of the three) observation used for the calculation of the derivative. We find that trigger probabilities calculated with spacings of up to $\sim$ 16 days provide similar predictions. However as the spacing is increased, predictions diverge at the large triggers. In particular, for the larger sampling rates, very large triggers are rarely observed because the sampling length is larger than or comparable to the event time-scale. The functions $F_{T}(P|T)$ are only useful for $P$ larger than the sampling rate. Note that in realistic situations the sampling rate will not be regular due to a variety of problems (weather etc.). However $F_{T}(P|T)$ predicts light-curve behaviour based upon the observation of a single derivative. A regular sampling rate over an extended period is therefore not essential.

\subsection{Triggering functions for different classes of HMEs}
\label{sec_classes}
\begin{figure*}
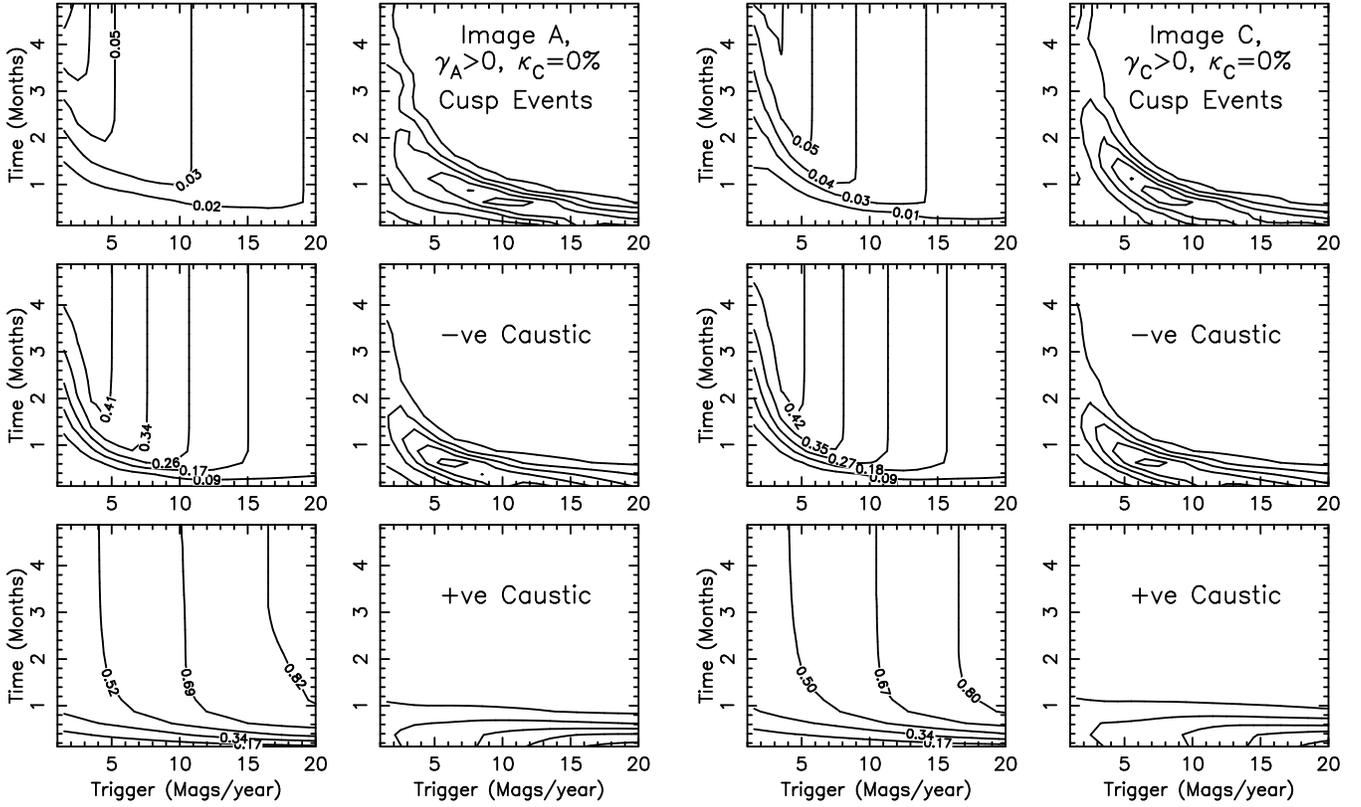

\vspace{115mm}
\includegraphics{fig8a.epsi}
\includegraphics{fig8b.epsi}
\caption{\label{contour_classes}Contour plots for $F_T(P|T)$ (left) and $\partial F_T(P|T)/\partial P$ (right) for images A and C. $F_T(P|T)$ and $\partial F_T(P|T)/\partial P$ are shown for 3 separate classes of events: cusp (top), caustic crossings with disappearing (centre) and appearing (bottom) critical images respectively. There is no smooth matter in these models, small photometric errors (SE) and $\gamma_A,\gamma_B>0$ are assumed.}
\end{figure*}

Due to the appearance ($+ve$ event) or disappearance ($-ve$ event) of a pair of critical images, caustic crossing HMEs are asymmetric in time. In particular, the light-curve remains at a higher level following the event if a new pair of critical microimages is created. In addition, the creation of critical images causes the light-curve to rise much more steeply than it falls. Where there is some knowledge about the type of an impending event (ie. whether the critical images are appearing or disappearing) the triggering function can be determined more accurately. A third possibility is that an HME may not involve the source crossing a caustic but be due to the source moving past a cusp. This type of event is not interesting for a spectroscopic measurements (eg. Wambsganss \& Paczynski 1991), and so forewarning would be useful. We have determined the functions  $F_{T_+}(P|T)$, $\partial F_{T_+}(P|T)/\partial P$;  $F_{T_-}(P|T)$, $\partial F_{T_-}(P|T)/\partial P$; and $F_{T_C}(P|T)$, $\partial F_{T_C}(P|T)/\partial P$ as before, looking for triggers that portend an event due to only appearing ($+ve$) or disappearing ($-ve$) critical images, or a cusp respectively within a given period. Note that
\begin{equation}
F_{T_+}(P|T)+F_{T_-}(P|T)+F_{T_C}(P|T)=F_{T}(P|T), 
\end{equation}
so at large $P$, $F_{T_+}$, $F_{T_-}$ and $F_{T_C}$ describe the relative fractions of different event types found by that trigger. 

 Fig. \ref{contour_classes} shows contours of $F_{T_+}(P|T)$, $\partial F_{T_+}(P|T)/\partial P$; $F_{T_-}(P|T)$, $\partial F_{T_-}(P|T)/\partial P$; and $F_{T_C}(P|T)$, $\partial F_{T_C}(P|T)/\partial P$. Because the rising side of a $-ve$ event (or a cusp event) is not as steep as that of a $+ve$ event, very large triggers are not likely to precede a $-ve$ event (or a cusp event). However, in the case where a large trigger is observed prior to a $-ve$ event, the delay $P$ is small. In contrast, a small trigger portends $+ve$ and $-ve$ events in approximately equal numbers. Similarly $\partial F_{T_-}(P|T)/\partial P$ has a maximum for an event arriving $\sim 3$ weeks after a trigger of $\sim$ 6 magnitudes per year, however $\partial F_{T_+}(P|T)/\partial T$ is larger at higher $T$ and smaller $P$. In addition, following a trigger $\ga 5$ magnitudes per year, $+ve$ caustic crossing events are more likely, and are expected sooner than $-ve$ caustic crossing events.

$F_{T_C}(P|T)$ and $\partial F_{T_C}(P|T)/\partial P$ behave similarly to the corresponding functions for $-ve$ caustic crossings. For smaller triggers, the delay until an event is larger. The similarity in shape of 
$F_{T_C}(P|T)$ and $F_{T_-}(P|T)$ will make distinguishing between a cusp and  $-ve$ caustic crossing difficult prior to the event peak. On the other hand $F_{T_C}(P|T)<F_{T_-}(P|T)$ and so cusp events are less common.

\subsection{Selecting a trigger}
\label{selecting}

\begin{figure}
\vspace{60mm}
\includegraphics{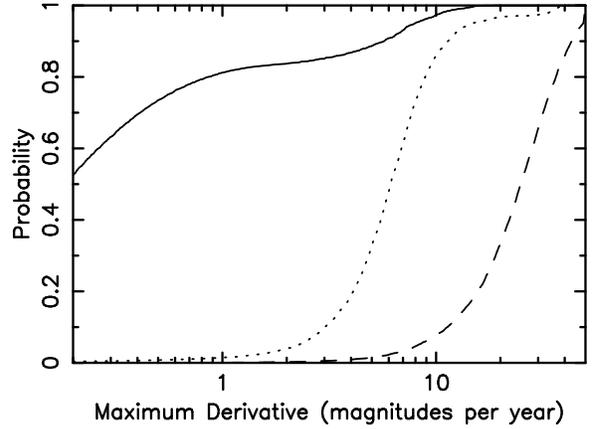}
\caption{\label{max_deriv}Plots of the probability for the maximum derivative preceding each a light-curve peak. Functions are given corresponding to derivatives that precede cusps (solid line), and caustic crossings with disappearing (dotted line) and appearing (dashed line) critical images. Only peaks larger than 0.1 magnitudes above the local minima on both sides are considered.  }
\end{figure}

While a derivative observed from monitoring data can be analysed as a potential trigger in real time it is useful to have a pre-conception of the optimal trigger. In particular, a program could demand that the trigger be larger than the maximum derivative expected for cusps. However, if this approach were taken, then some $-ve$ caustic events may peak before a TOO was triggered. A large trigger will always yield the desired caustic crossing, however HMEs may be missed before such a trigger is observed (and HMEs are rare). On the other hand if the TOO is triggered on every low derivative then the success rate will be lower, but the observation of the next HME is guaranteed. A balance must be struck between the quantity of available TOO observing time, and how long one is willing to wait to obtain the observations.

We have calculated the cumulative probability of observing a maximum light-curve derivative prior to the peak maximum: 
\begin{eqnarray}
\nonumber
&&\hspace{-7mm}P_{\dot{M}}(\dot{M}<\dot{M}_o)=\\
\nonumber&&\hspace{-3mm}\int dm\int dS\int dv_{eff}\,\,\left( p_{s}\left(S|\langle m\rangle,v_{eff}\right)\,p_{m}\left(\langle m\rangle\right)\right.\\
&&\hspace{5mm}\times\left.p_{v}\left(v_{eff}|\langle m\rangle\right)P_{\dot{M}}\left(\dot{M}<\dot{M}_o|S,\langle m\rangle,v_{eff}\right)\right)
\end{eqnarray}
using a sampling rate of 1-week. We have only presented the case of models with no smooth matter, small photometric errors (SE), $\gamma_A,\gamma_B>0$ and no intrinsic source variability. We have looked at the probability $P$ for cusps, $P_{\dot{M}_C}(\dot{M}<\dot{M}_o)$; $-ve$ caustic crossings, $P_{\dot{M}_-}(\dot{M}<\dot{M}_o)$, and $+ve$ caustic crossings, $P_{\dot{M}_+}(\dot{M}<\dot{M}_o)$. 
Fig. \ref{max_deriv} shows these functions; solid, dotted and dashed lines correspond to $P_{\dot{M}_C}(\dot{M}<\dot{M}_o)$, $P_{\dot{M}_-}(\dot{M}<\dot{M}_o)$, and $P_{\dot{M}_+}(\dot{M}<\dot{M}_o)$ respectively.

Fig. \ref{max_deriv} shows that a trigger of $\sim5$ magnitudes per year will successfully trigger observations for most $-ve$ caustic crossings, all $+ve$ caustic crossings, and miss $\sim 90\%$ of cusp events. The source size has been estimated from the 1988 event in R-band. The trigger should therefore be from observations made in R-band since then the triggering function is more accurately determined. Also, accretion disc models predict that the source should be smaller in higher frequency bands: monitoring and triggering at a higher frequencies therefore provides more rapid variations, but gives less warning of the impending event.

\subsection{An alternative approach}
\label{sec_alternative}
Recall that the method to determine $F$ steps along the light-curve until a derivative is located that is greater than $T$, the search is then resumed after the subsequent event. The probabilities therefore refer to the fractions of events of each type that are located on a light-curve and have a derivative greater than $T$ on their leading side (integrated over effective transverse velocity). This calculation is most useful when an immediate history of image brightnesses is available, so that it is known that the current derivative is the first above $T$.

Alternatively, the a-priori probability of observing an event following a light-curve trigger $T\pm\Delta T$ can be computed by finding the derivative and separation of the derivative and corresponding event at all (equally spaced) points on the light-curve. This calculation does not take into account the recent history of the light-curve, and should be applied to an isolated observation (eg the first point of the observing season. The resulting $F_{T_+}(P|T\pm \Delta T)$, $F_{T_-}(P|T\pm \Delta T)$ and $F_{T_C}(P|T\pm \Delta T)$ describe the likely-hood of the class of event following the trigger in the absence of previous light-curve information rather than the relative numbers of events of each type that are located on a light-curve.

\section{A Mock Target of Opportunity Experiment}

In this section we combine high resolution microlensing magnification grids with a thermal accretion disc source to produce model microlensed light-curves in 11 bands. We then produce example target of opportunity spectroscopic observations for hypothetical light-curve triggers. By looking at the microlensing induced change in spectral slope we estimate the chance of success for a three observation TOO program. The three spectra are taken at intervals fixed at the time of triggering such that they are likely to observe the light-curve peak (ie. according to $F(P|T)$). We emphasise that this section does not predict the colour changes that will be measured, rather it demonstrates that a physically motivated source with a spectrum that is radius dependent will produce colour change that could be detected by target of opportunity observations.

\subsection{The model accretion disc}

We use a standard thin accretion disc with constant accretion rate and no advection of heat. The energy generation per unit area as a function of radius is given by 
\begin{equation}
Q=\frac{3}{4\pi}\frac{GM_{BH}\dot{\cal M}}{r^{3}}R_R(r).
\end{equation}
$M_{BH}$ is the black hole mass, $\dot{\cal M}$ is the accretion rate, $r$ the radius and $R_R$ is a correction factor. $R_R$ combines outward advection of energy associated with the angular momentum flux and relativistic effects (Page \& Thorne 1974, Krolik 1998). To calculate the flux generation at the surface of an accretion disc, the relativistic effects of Doppler beaming of radiation, gravitational redshifts, and bending of photon trajectories must be considered. These are included through the calculation of a relativistic transfer function (Cunningham 1975). The following assumptions are made to compute the transfer function: $i$) the accretion disc is thin (disc height much smaller than disc radius), $ii$) the gas follows prograde circular orbits outside the marginally stable radius $r_{ms}$, $iii$) the disc is flat and lies in the equatorial plane of the black hole and $iv$) the gas emits isotropically in its rest frame. Note that this model does not emit inside $r_{ms}$. The disc model is discussed in more detail in Agol \& Krolik (1999).

\subsection{The model accretion disc light curves}

\begin{figure}
\vspace{60mm}
\includegraphics{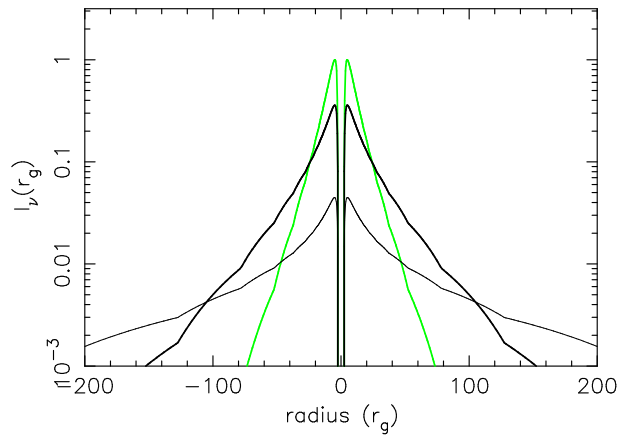}
\caption{\label{sourcees}Cross-sections of the intensity profiles in the U (light line), R (thick dark line) and K (thin dark line) for the thin accretion disc source. $I_{\nu}$ is the intensity per unit frequency on the surface of the accretion-disc, normalised to the maximum value in U-band. The profile is shown for a disc viewed face on, hence the deficit of flux at the centre.}
\end{figure}

\begin{table}
\begin{center}
\caption{\label{freqs} The frequencies at the centre of the bands for which source intensity profiles and light-curves have been calculated.}
\begin{tabular}{|c|c|c|}
\hline
$\nu$ $(Hz)$ &  band \\ \hline\hline
5.50$\times 10^{15}$      &  F160AW \\ 
3.90$\times 10^{15}$      &  F218AW \\
2.77$\times 10^{15}$      &  U (F300W) \\
2.06$\times 10^{15}$      &  U'(F380W) \\
1.73$\times 10^{15}$      &  B \\
1.48$\times 10^{15}$      &  V \\
1.20$\times 10^{15}$      &  R \\
1.01$\times 10^{15}$     &  I \\
6.47$\times 10^{14}$     &  J \\
4.90$\times 10^{14}$     &  H \\
3.63$\times 10^{14}$      &  K \\ \hline
\end{tabular}
\end{center}
\end{table}

Several attempts have been made to understand the accretion disc in Q2237+0305 by comparing model light curves with observed microlensing events (eg. Rauch \& Blandford 1991; Jaroszynski et al. 1992; Jaroszynski \& Marck 1994; Czerny et al. 1994). The approach of these papers is to discuss the luminosity of the quasar considering the predicted magnification of the macro-images as well as the limits on source size imposed by microlensing observations. In this paper we compute light-curves for the accretion disc in 11 frequency bands (see table \ref{freqs}): The quoted frequency is the band centre in the quasar rest frame, the first 4 filters are HST filters. Fig. \ref{sourcees} shows the cross-sections of the intensity profiles in the U (light line), R (thick dark line) and K (thin dark line). The source size ($S$) must be interpreted in terms of its intensity profile. We define the source size such that the event length produced by the source is equivalent to that produced by a top-hat profile of diameter $S$. We have used the example of a black hole mass of $10^{8}M_{\odot}$ which at 1.2$\times 10^{15}\,Hz$ (R-band) corresponds to a source radius of $5\times 10^{14}\,cm$. This model suffers the difficulties that were pointed out in the case of a thermal accretion disc by Rauch \& Blandford (1991), namely that it severely underestimates the observed flux. However our focus here is to combine a detailed accretion disc model with high resolution microlensing models to simulate microlensing induced spectral change during a caustic crossing.

The source profile is computed over a nested grid of cells (with side length 500$r_g$) down to a resolution of 0.5 $r_g$ in each of the 11 bands. The grid contains 4864 cells which lie on 192 parallel source tracks. The light-curves were computed using the method described by Wyithe \& Webster (1999), and computed at a resolution of 1.5$\times 10^{-4}\eta_o$ along the source line. At an effective galactic transverse velocity of 400$\,km\,sec^{-1}$, this corresponds to a temporal resolution of $\sim 0.5$ days. 
 
\subsection{Examples of model HME observations}

\begin{figure*}
\vspace{70mm}
\includegraphics{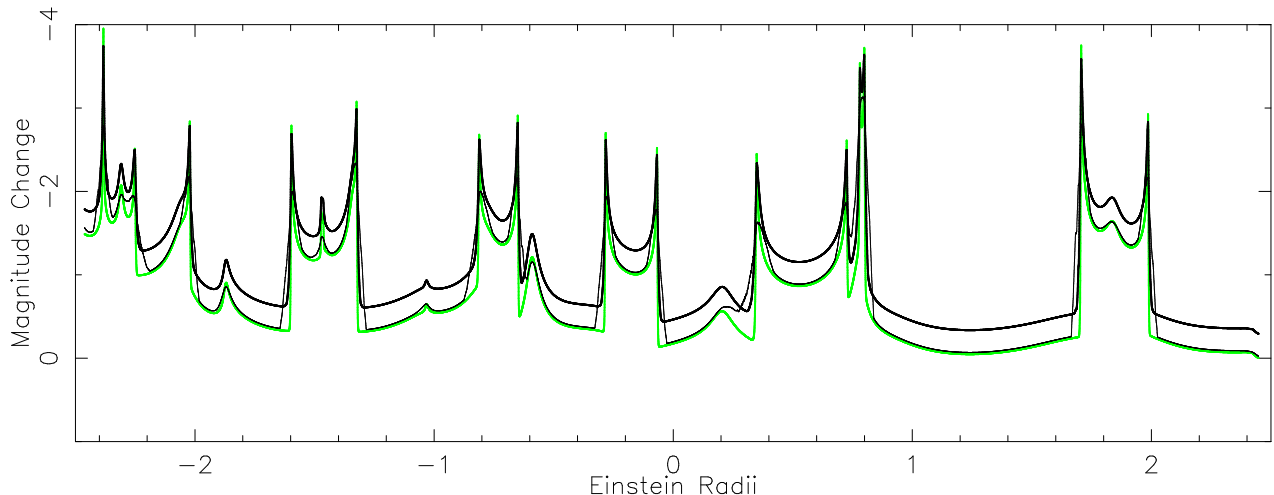}
\caption{\label{lcurves}Model light-curves of image A for the U-band (light line), R-band (thick dark line) and K-band (thin dark line) sources. The microlensing model had a mean microlens mass of $\langle m\rangle=0.1M_{\odot}$, and the source black-hole mass was $M_{BH}=10^8M_{\odot}$.}
\end{figure*}

\begin{figure*}
\vspace{220mm}
\includegraphics{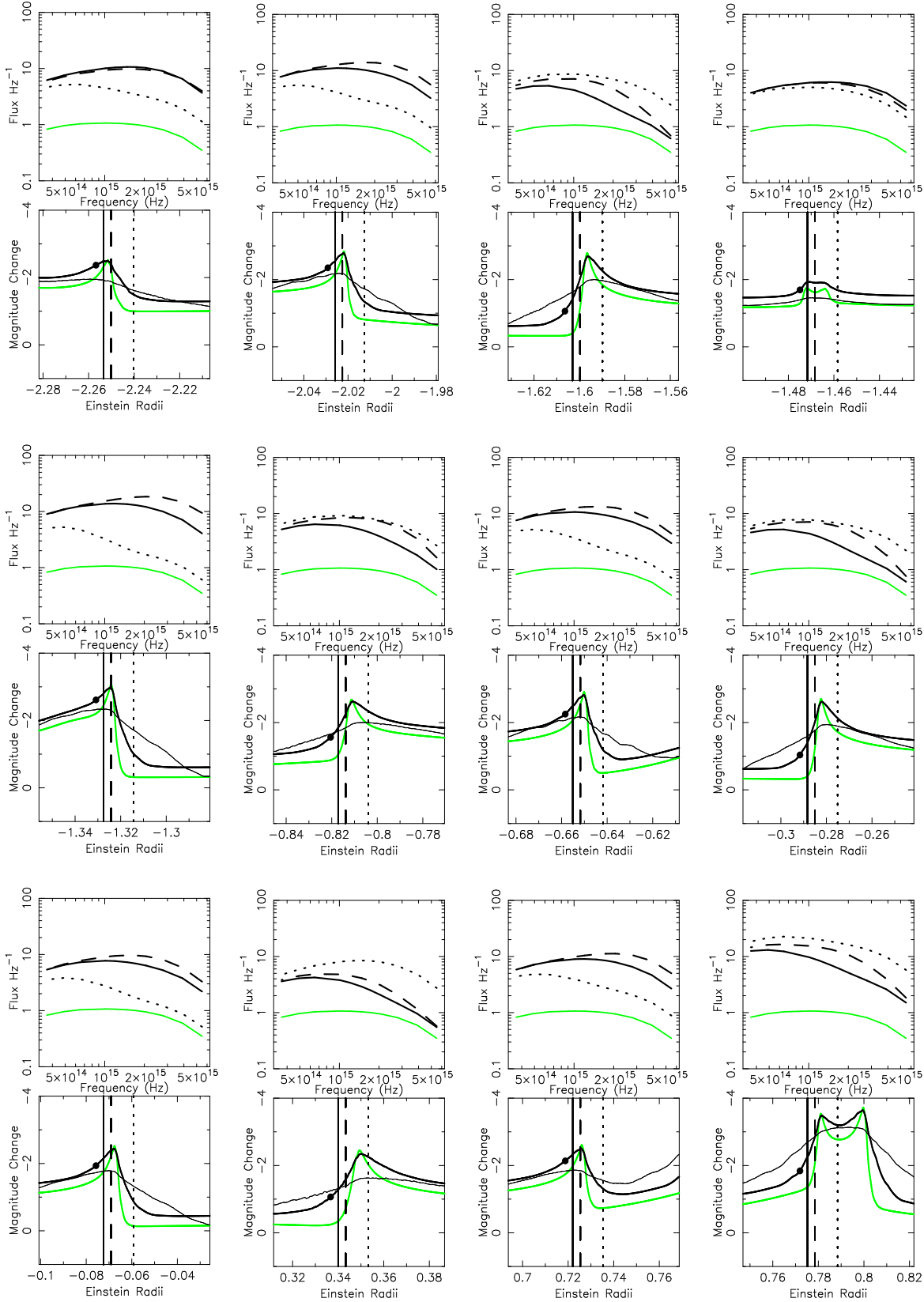}
\caption{\label{trigger_5} Examples spectrophotometric simulations that result from a trigger of $T=5$ magnitudes per year from the R-band light-curve shown in Fig. \ref{lcurves}. In each case, the three spectra (flux per unit frequency vs. frequency) are shown in the upper panel. An unlensed spectrum is shown for comparison (light line). The lower panels show close-ups of the event in the U, R and K bands, the vertical lines show the position of the three observations on the light-curve and have line-styles that correspond with the spectra in the upper panel. The position of the trigger is denoted on the R-band light-curve by a dot.}
\end{figure*}

\begin{figure}
\vspace{220mm}
\includegraphics{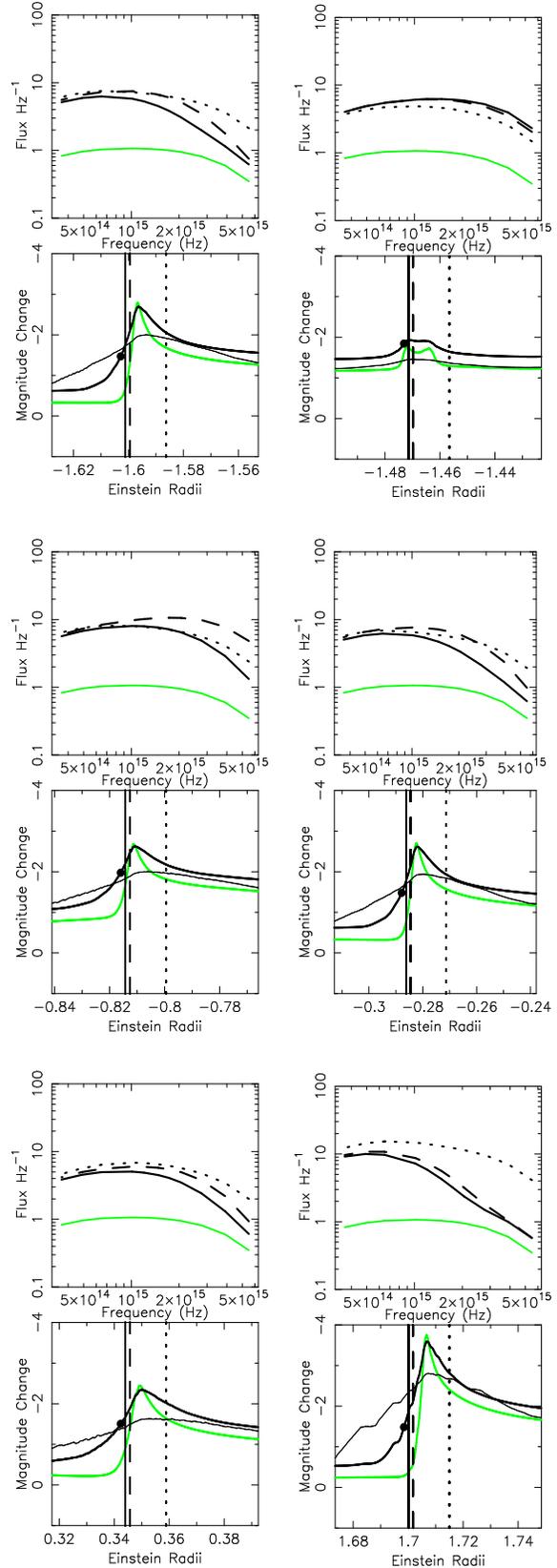}
\caption{\label{trigger_15}Examples spectrophotometric simulations that result from a trigger of $T=15$ magnitudes per year from the R-band light-curve shown in Fig. \ref{lcurves}. Presentation as per Fig.~\ref{trigger_5}} 
\end{figure}

\begin{figure*}
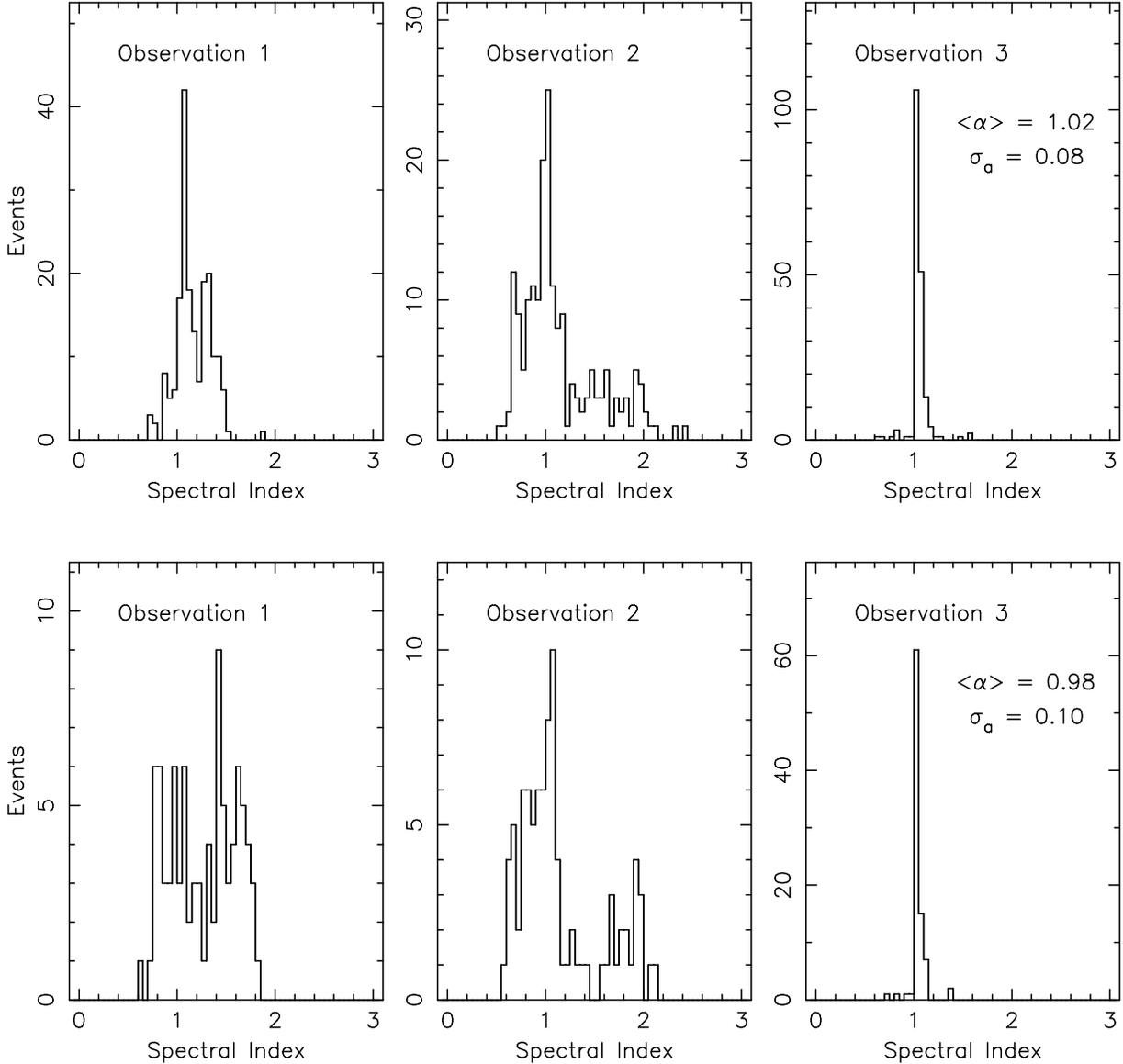

\vspace{160mm}
\includegraphics{fig14a.epsi}
\includegraphics{fig14b.epsi}
\caption{\label{spec_index_15}Histograms of the microlensed spectral slope between HST bands F380W (2.06$\times 10^{15}$Hz) and F160AW (5.50$\times 10^{15}$Hz). Histograms for three observations are shown. Top: 2, 4 and 10 weeks following an R-band trigger of 5 magnitudes per year. Bottom: 1, 2 and 10 weeks following an R-band trigger of 15 magnitudes per year.}
\end{figure*}

\begin{figure*}
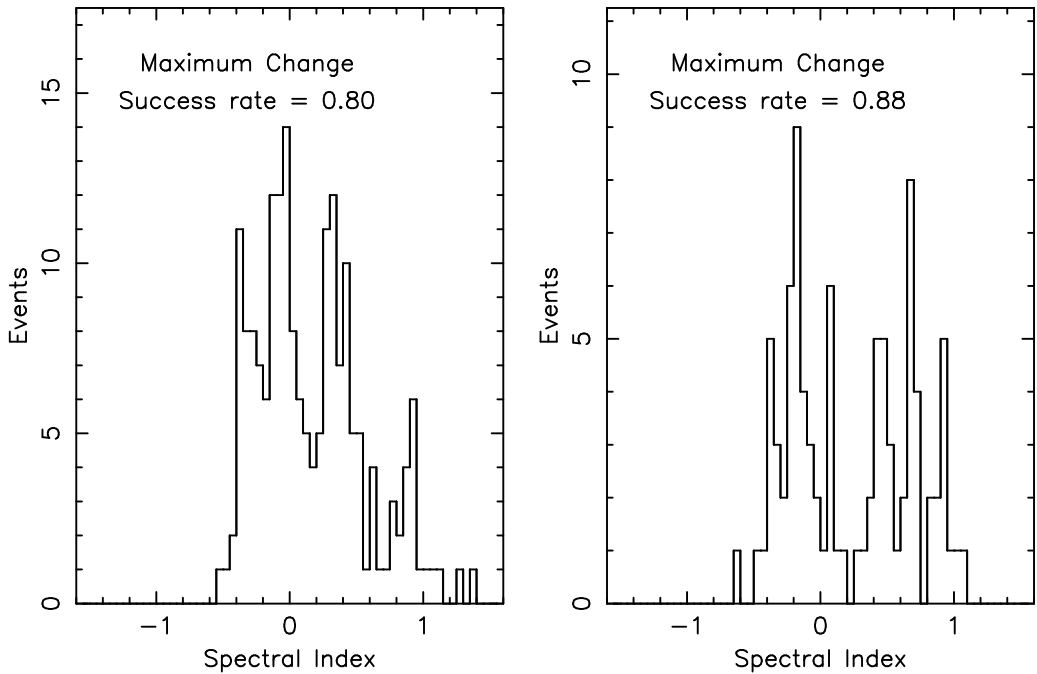

\vspace{100mm}
\includegraphics{fig15a.epsi}
\includegraphics{fig15b.epsi}
\caption{\label{spec_index_diff}Histograms of the larger difference between the microlensed spectral slope of observation 1 or 2 and observation 3. the spectral slope was calculated between between HST bands F380W (2.06$\times 10^{15}$Hz) and F160AW (5.50$\times 10^{15}$Hz) following an R-band trigger of Left: 5 magnitudes per year, and Right: 15 magnitudes per year.}
\end{figure*}

\begin{figure*}
\vspace{75mm}
\includegraphics{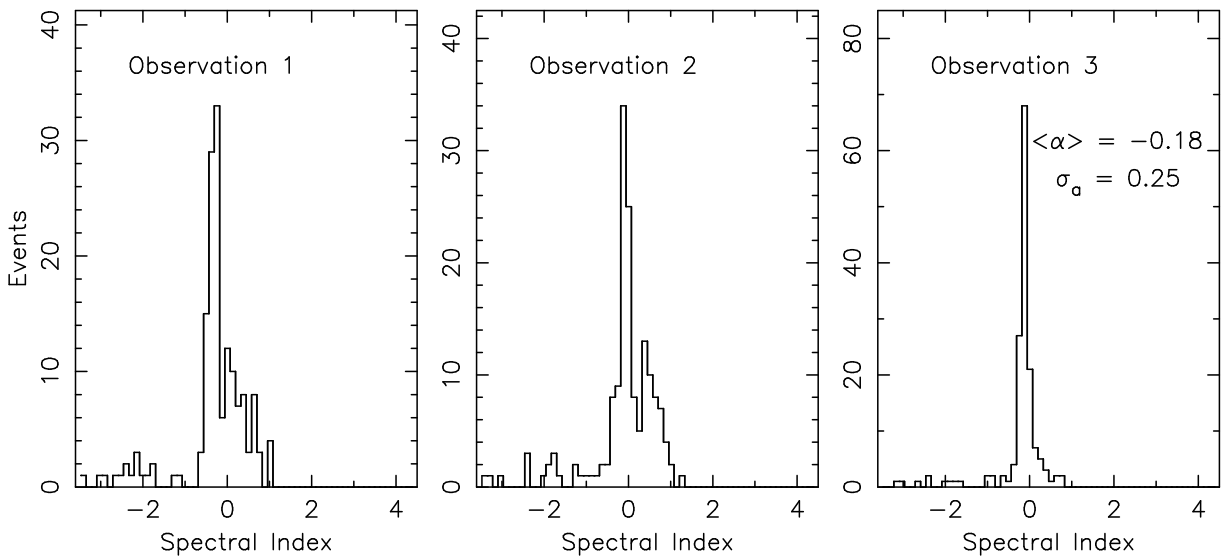}
\caption{\label{spec_index_5_low_freq}Histograms of the microlensed spectral slope between K and B bands (3.63$\times 10^{14}$Hz) and (1.73$\times 10^{15}$Hz) for the three observations at 2, 10 and 30 weeks following an R-band trigger of 5 magnitudes per year.}
\end{figure*}

We calculate the microlensed light-curves of an accretion disc source having a central black-hole mass of $10^{8}M_{\odot}$ ($S_R\sim 10^{15}\,cm$ in R-band). The mean microlens mass in the simulation was $\langle m\rangle =0.1M_{\odot}$ and the microlensing parameters corresponded to image A ($\gamma_{A}=0.4$). The results for $S$ and $\langle m\rangle$ of WWTM00 suggest that these parameters are respectively near the upper and lower limits expected for Q2237+0305. The colour changes found are therefore conservative for this class of disc model. Example light-curves with the aforementioned parameters are presented in Fig. \ref{lcurves}. Three curves are shown corresponding to the source profiles in Fig. \ref{sourcees}. The light-curves show magnitude change plus an offset equal to the difference in unlensed flux.

Ensembles of mock experiments are produced for two hypothetical triggers: $T=5$ and $T=15$ magnitudes per year. Below we simulate a TOO comprising 3 spectroscopic observations taken at intervals designated at the time of the trigger. The first two observations are scheduled according to when the triggering function $\partial F_{T}(T,P)/\partial P$ predicts an HME: $2-4$ weeks following the small trigger and $1-2$ weeks following the large trigger. A third observation is made 10 weeks following the trigger to provide a comparison with the spectrum of a (hopefully) non-differentially magnified source.

 We assume a likely effective galactic transverse velocity of $400\,km\,sec^{-1}$, and trigger observations from the R-band light curve in Fig. \ref{lcurves}. Examples of the spectrophotometric simulations are presented in Figs. \ref{trigger_5} and \ref{trigger_15}. In each case, four spectra are shown in the upper panel, three microlensed spectra as well as the unlensed spectrum (light-line) for comparison. The lower panels show close-ups of the events in the U, R and K bands, the vertical lines show the position of the three observations on the light-curve and have line-styles that correspond with the spectra in the upper panel. The position of the trigger is denoted on the R-band light-curve by a dot.

In the example shown the black hole mass of $10^8M_{\odot}$ produces a larger source than microlensing observations predict is most likely. The triggering function, which is calculated using the most likely source size therefore underestimates the delay between the trigger and event peak in several cases (particularly for the larger trigger) because larger sources experience HME scale derivatives earlier than small sources. However, the spectral change is at an observable level in nearly all cases. The size of the spectral change observed during the HME is qualitatively dependent on whether the event is a $+ve$ or a $-ve$ caustic crossing. In the $-ve$ case, the first two observations are taken during the event. The final observation is made following the disappearance of the critical images at a time when the source has little differential magnification. During the first two observations, the higher frequencies (emitted from smaller regions) are preferentially magnified, and so the spectrum is stronger at the higher frequencies. In contrast, the $+ve$ events have initial observations that are taken before the event has peaked. There are no critical images of the inner regions and hence they are not as strongly magnified. The final observation has critical images of the entire source, and so an equal magnification in all bands. In both cases, the spectra plus a well sampled light-curve in a single band will allow determination of the relative sizes of the emitting regions at all frequencies which are in contact with the caustic for the first two observations. The trigger of $T=5$ magnitudes per year finds $+ve$ and $-ve$ events in approximately equal numbers. However, the trigger of $T=15$ magnitudes per year finds only $+ve$ events because $-ve$ events do not reach such large light-curve gradients before the event peak.

To quantify the effect of the model TOO experiment, we calculate the distribution of spectral slopes for each of the three observations in an ensemble of model TOOs. The triggering function was applied to $50\eta_o$ ($\sim 500$ years at $400\, km\,sec^{-1}$) of light-curves for the source and microlensing model discussed above, and continuum spectra calculated for three observations following the trigger. The 5 magnitudes per year trigger precedes 159 HMEs in these light-curves. However the larger trigger (15 magnitudes per year) only precedes 80 HMEs since such a large light-curve derivative is unlikely to precede $-ve$ caustic crossings or cusp events. We have calculated the change in spectral slope $\alpha$ (over the wavelength range $F160AW - U'$) for each event, which is defined as:
\begin{equation}
\alpha\equiv\frac{log\left(F_{\nu}(F160AW)\right)-log\left(F_{\nu}(U')\right)}{log\left(\nu(F160AW)\right)-log\left(\nu(U')\right)}
\end{equation}
Fig. \ref{spec_index_15} shows the distributions of spectral slope for the three observations following triggers of $T=5$ (upper panels) and $T=15$ (lower panels) magnitudes per year. The results obtained for the two triggers are reassuringly similar. There is a large spread about the unlensed $\alpha=1$ in the spectral indices obtained from the first two observations. As mentioned, differential magnification from microlensing can produce both an increase and a decrease in the spectral slope. This is reflected in the distributions which are approximately symmetric about $\alpha=1$. The third observation at 10 weeks is intended as a control and Fig. \ref{spec_index_15} shows that the source is rarely subject to strong differential magnification at this time. The range of spectral indices for this observation is small, and its mean is consistent with the unlensed value of $\alpha\sim1$.

Fig. \ref{spec_index_diff} shows histograms of the larger difference (in absolute value) between observation 1/2 and observation 3 following triggers $T =5$ and $T =15$ magnitudes per year. 
For a successful measurement of microlensing induced change in spectral slope to be made, a change in $\alpha$ will need to be observed both at a level above observational uncertainty, and the expected spread of values for observation 3.  Comparison of Figs. \ref{spec_index_15} and \ref{spec_index_diff} shows that this is true in most cases. To quantify the chance of success for the experiment we define a TOO program to be successful when one of the initial two observations lies outside the variance of the third observation (computed from the entire ensemble). We find that this definition yields encouraging success rates of $\sim 80\%$ and $\sim 88\%$ for triggers of $T =5$ and $T =15$ magnitudes per year respectively. Of course if more than three observations were available, the chances of success would be enhanced. Unsuccessful TOOs may be due to triggering on a cusp event (more likely for $T=5$, hence the lower success rate) or to an unusual caustic crossing whose peak lay outside the expected range.

The above calculation considers only the effect of microlensing on the spectral slope at high-frequency. While this is the regime where microlensing is expected to have the largest effect it is also interesting to see if similar information  can be obtained at longer wavelengths, specifically from the spectral slope between the $K$ and $B$ bands:
\begin{equation}
\alpha\equiv\frac{log\left(F_{\nu}(K)\right)-log\left(F_{\nu}(B)\right)}{log\left(\nu(K)\right)-log\left(\nu(B)\right)}.
\end{equation}
 In this case the source is much larger and so the control observation must be taken at a much later time. Fig. \ref{spec_index_5_low_freq} shows the histogram of spectral indices for three observations at 2, 10 and 30 weeks following a trigger of $T=5$ magnitudes per year. The control observation has a large spread (3 times that for the high frequency index), indicating that even after 30 weeks, there is still some differential microlensing due to the caustic. The TOO experiment is much more likely to be successful at higher frequencies.

\section{conclusion} 

We have calculated the probability (triggering function) of observing a caustic crossing within a given period following the observation of a rise (trigger) in an image light-curve of Q2237+0305. We find that there is a well defined mode for this distribution, that the mode occurs earlier, and that the distribution peak becomes narrower for observations of larger light-curve derivatives. Thus, the more rapid the observed rise in monitoring data, the sooner a caustic crossing is expected, and the more accurate its predicted arrival time will be. Predictions of caustic crossings are aided by knowledge of whether a pending event will be due to a pair of appearing or disappearing critical images. In particular, an event that is due to appearing critical images is expected sooner than one due to disappearing critical images. The probabilities for cusp events are similar to those for events due to disappearing critical images, making distinguishing these difficult before the event event has reached its maximum. However cusp events are far rarer than caustic crossings.

We have checked the effect on our triggering function of different combinations of assumptions for the size of photometric error in the monitoring data, the fraction of smooth matter content and the direction of the galactic transverse velocity. We find little systematic dependence on these quantities. We have placed limits on the intrinsic variation in Q2237+0305 and shown that (except for the smaller triggers) the predictions should not be effected by intrinsic source variation. A systematic underestimation (the largest possible) of $\times 2$ in our previous estimation of source size would produce an underestimate in the arrival time. The underestimation is most serious for large triggers preceding a caustic crossing high magnification event. Predictions from smaller derivatives are less affected.

Computation of the distributions of the separation between a trigger and subsequent high magnification event for different sampling rates determines the appropriate sampling rate for the prediction of caustic crossings from regular monitoring. We find that the triggering function is insensitive to sampling rate for sampling rates of $\sim 2$ weeks or less. If caustic crossing probabilities are calculated from monitoring data with a higher sampling rate, the derivative may be calculated over a longer baseline to reduce the photometric uncertainty.

We have computed the light-curves of a thermal accretion disc about a central black-hole of $10^8M_{\odot}$ in 11 bands. We look for hypothetical triggers on the model R-band light-curves of $T=5$ and $15$ magnitudes per year and compute two model spectra at times most likely to surround the peak (determined from the triggering function). A third spectrum is computed 10 weeks following the trigger and is intended as a standard (not differentially magnified). We find that the third observation is subject to very little differential magnification and the computed spectrum has a slope $\alpha$ (between $2.06\times10^{15}\,Hz$ and $5.50\times10^{15}\,Hz$) that is barely changed from that of the unlensed source ($\alpha\sim1$). However the initial two spectra have spectral slopes that differ from that of the unlensed source by between -0.5 and 1.

We have calculated the chance of success of a target of opportunity program that contains 3 observations. The success rate is defined as the fraction of an ensemble of mock target of opportunity observations that have one of the initial two observations with a spectral slope outside the variance in the third (calculated over the entire ensemble). We find that the success rate should be high, $\ga80\%$, which demonstrates that a predefined triggering and TOO observational strategy can achieve the goal of observing spectrophotometric change during a microlensing event in Q2237+0305.

\section{acknowledgements}
This work was supported by NSF grant AST98-02802. JSBW acknowledges the support of an Australian Postgraduate Award and a Melbourne University Postgraduate Overseas Research Experience Award.

\label{lastpage}

\end{document}